\documentclass[bibyear]{aa}  
\usepackage{graphicx}
\usepackage{txfonts}
\begin{document}

   \title{\textit{Kepler} super-flare stars: what are they?}

   \subtitle{}

   \author{
     R. Wichmann\inst{1} 
     \and
     B. Fuhrmeister\inst{1}
     \and
     U. Wolter\inst{1}
     \and
     E. Nagel\inst{1}
    }
   
   \institute{
     Hamburger Sternwarte,
     Gojenbergsweg 112, 21029 Hamburg\\
     \email{rwichmann@hs.uni-hamburg.de}
   }

   \date{Received date; Accepted date}

% \abstract{}{}{}{}{} 
% 5 {} token are mandatory
 
   \abstract{
     The {\it Kepler} mission has led to the serendipitous discovery 
     of a significant number of `super flares' - white light 
     flares with energies between 10$^{33}$\,erg and 10$^{36}$\,erg -
     on solar-type stars. It has been speculated that these could 
     be `freak' 
     events that might happen on the Sun, too.
     We have started a programme to study the nature of the stars on 
     which these 
     super flares have been observed. Here we present high-resolution 
     spectroscopy of 11 of these stars and discuss our results.
     We find that
     several of these stars are very young, fast-rotating stars where high
     levels of stellar activity can be expected, but 
     for some other stars we do not find a straightforward explanation 
     for the occurrence of super flares.
   }


   \maketitle
%
%________________________________________________________________

\section{Introduction}

   The NASA {\it Kepler} space observatory has collected high-precision
   photometric measurements at a cadence of 29.4\,min for $\sim$\,156~000 
   stars (Jenkins et al. \cite{jenkins}). While the primary goal of the
   {\it Kepler} mission has been to detect extrasolar planets, the
   large amount of photometric data has enabled serendipitous observation
   of large numbers of rare yet highly energetic `super flares' on solar-type
   stars (Maehara et al. \cite{maehara} and Shibayama et al. \cite{shibayama}).

   The super flares detected by these authors are white-light flares 
   with bolometric 
   energies of 
   $\geq 10^{33}$\,erg that typically reach 0.1 to 1 per cent of the stellar
   luminosity, and they last for a few hours. For comparison, the largest known
   solar flare - the Carrington event on Sept 1, 1859 
   (Carrington \cite{carrington}) - had an estimated energy of the order of
   $\geq 10^{32}$\,erg (Tsurutani et al. \cite{tsurutani}).

   It is intriguing that flares of such magnitude can occur on solar-type stars,
   in particular because some of these events took place on slowly
   rotating stars ($> 10$\,days rotation period). Maehara et al. 
   (\cite{maehara}) find that the frequency of super flares on slow rotators 
   is only 1/20 of that for fast rotators, so there is clearly a strong
   tendency towards fast-rotating, hence statistically young stars. Still,
   one may ask whether the Sun itself is also capable of producing such flares
   (albeit rarely) or whether the stars showing such events are in some 
   way special. So far only a few spectroscopic studies of {\it Kepler} 
   super-flare stars have been carried out. Notsu et. al (\cite{notsu}) 
   have studied the G-type star KIC 6934317, which exhibits a low lithium 
   abundance combined with 
   fast rotation (though apparently observed nearly pole on). It has 
   a high rate of super flares, but does  not seem to be particularly young.
   Nogami et al. (\cite{nogami}) have studied the slowly rotating stars
   KIC 9766237 and KIC 9944137 (rotation periods 21.8d and 25.3d, respectively)
   and report that both stars are similar to the Sun with respect to 
   metallicity, temperature, and surface gravity. They do not find any hint
   of binarity, although they cannot firmly rule out a low-mass companion.

   We aim to study the nature of the stars on which super flares have been 
   observed and have started a programme to carry out high-resolution
   spectroscopic observations of selected stars from the list of
   Maehara et al. (\cite{maehara}).
   In particular, we are investigating whether these stars show other
   indicators that are usually associated with high levels of activity, 
   such as fast rotation or high lithium $\lambda$\,6707 equivalent 
   width (EW), which indicate stellar youth, or rapid line shifts showing 
   that the star is in a close binary system.

   The paper is structured as follows. In Sect. 2 we describe the 
   observations and data collected.
   Section 3 explains our general methods and results, and in Sect. 4 we
   discuss the individual stars of our sample.

%
%______________________________________________________________
%

\section{Target selection and observations}

   Our programme started as a backup programme for the Calar Alto 
   2.2\,m telescope with the CAFE spectrograph, which meant that
   the primary
   selection criteria was that the stars had to be bright enough 
   for observations with a 2.2\,m telescope and had to be
   observable at the time of the original run.
   Target stars were observed multiple times in order to detect
   eventual radial velocity variations caused by multiplicity. We aimed
   for a S/N of about 50 when choosing exposure times, but owing to
   bad weather  we could not reach this goal for all observed
   stars.

   The first observing run took place from 
   Aug 17 to Aug 21, 2012 at the Calar Alto 2.2\,m telescope with 
   the CAFE echelle spectrograph (R $\simeq 65000$, spectral range 390-960\,nm,
   88 orders).
   During the first three nights, the weather was poor, while the last
   two nights had very good weather conditions. Table \ref{table:1} lists
   the stars observed in this run, along with stellar parameters from the
   literature. The values marked KIC are taken from Brown et al. (\cite{brown}),
   the values marked SDSS are from Pinsonneault et al. (\cite{pinsonneault})
   derived with the SDSS method. Rotation periods were derived 
   using the periodogram tool of the NASA exoplanet 
   archive\footnote{http://exoplanetarchive.ipac.caltech.edu/index.html}.

   For KIC 11390058 some episodes of the light curve do not show a stable 
   rotation pattern, as shown in Fig. \ref{kepler_lightcurves}. 
   However, 
   many light-curve intervals clearly yield a rotation period of 
   approximately 12 days, and we use the value of 11.9 days given in the 
   supplement of Maehara et al. (\cite{maehara}).

   For KIC 11972298 some light curve intervals 
   suggest a period of approx. eight days. 
   However, this seems highly unlikely in view of all available Kepler 
   quarters. As for KIC 11390058, many light curve intervals 
   clearly indicate a 
   rotation period of about 15 days, and again we use the 
   Maehara et al. (\cite{maehara}) value 
   of 15.5 days. We note that Shibayama et al. \cite{shibayama} prefer the
   shorter period and list 7.7 days for this star.

   We had a second observing run from May 28 to May 31, 2013. 
   Unfortunately, the S/N of most spectra of this run turned out to be too low
   to perform our parameter analysis.
   The three stars with usable data are listed in Table \ref{table:1}. 
   KIC\,7264976 has been 
   observed in 2012 and 2013, but the 2013 data only suffice to 
   determine radial velocities.
   Data reduction was performed using standard IRAF 
   \footnote{
     IRAF is distributed by the National Optical Astronomy 
     Observatories,
     which are operated by the Association of Universities for Research
     in Astronomy, Inc., under cooperative agreement with the National
     Science Foundation.
   }
   tasks for bias correction,
   flat fielding, order extraction, and wavelength calibration.
   
%______________________________________________________________
%
%_____________________________________________________________
%                                             Simple A&A Table
%_____________________________________________________________
%
\begin{table*}
\caption{
  Stars observed and their stellar parameters from the literature.
  KIC\,9653110, KIC\,11390058, and KIC\,11972298 were observed in the 2013 run.
  Periods for KIC\,11390058 and KIC\,11972298 are from Maehara et al. (\cite{maehara}).
}             % title of Table
\label{table:1}      % is used to refer this table in the text
\centering                          % used for centering table
\begin{tabular}{rrrrrrrr}        % centered columns (4 columns)
\hline\hline                 % inserts double horizontal lines
KIC & RA & Dec & $T_{\mathrm{eff}}$ KIC & log\,$g$ KIC & $T_{\mathrm{eff}}$ SDSS &  P \\  % table heading 
    & [h\,m\,s]   &[$^{\circ}$\, \arcmin \, \arcsec]    & [K]   &   & [K] &          [days]\\
\hline                        % inserts single horizontal line
   3626094 & 18\,59\,14.9 & +38\,45\,46 & 5835 & 4.3 &   6156 & 0.72  \\ % 3.766 
   4742436 & 19\,21\,49.5 & +39\,50\,07 & 5628 & 4.2 &   5980 & 2.3   \\ % 3.750
   4831454 & 19\,21\,58.0 & +39\,59\,54 & 5298 & 4.6 &   5536 & 5.2   \\ % 3.724
   7264976 & 19\,02\,36.1 & +42\,48\,31 & 5184 & 4.1 &   5400 & 12.7  \\ % 3.715
   8479655 & 18\,57\,43.2 & +44\,35\,55 & 5126 & 4.6 &   5415 & 19.3  \\ % 3.710
   9653110 & 19\,33\,48.3 & +46\,21\,54 & 5223 & 4.4 &   5474 & 3.2   \\ % 3.718
  11073910 & 19\,04\,27.7 & +48\,36\,55 & 5381 & 4.6 &   5640 & 5.5   \\ % 3.731
  11390058 & 18\,56\,45.9 & +49\,17\,30 & 5785 & 4.3 &   6086 & {\it 11.9}      \\
  11610797 & 19\,27\,36.7 & +49\,40\,14 & 5865 & 4.5 &   6090 & 1.6   \\ % 3.768
  11764567 & 19\,30\,33.6 & +49\,56\,04 & 5238 & 4.4 &   5480 & 20.5  \\ % 3.719
  11972298 & 19\,44\,23.3 & +50\,22\,12 & 5498 & 4.4 &   5741 & {\it 15.5}      \\
\hline                                   %inserts single line
\end{tabular}
\end{table*}

%
%______________________________________________________________
%

\section{Results}

\subsection{Determination of stellar parameters}

We determined the stellar parameters using two different methods. 
First, we utilised 'spectroscopy made easy' (SME, Valenti \& Piskunov \cite{valenti96}), 
and second we fitted
PHOENIX model spectra (Husser \cite{Husser}) to our observed spectra. For both
methods we used averaged spectra of our objects to improve the S/N.

\begin{figure*}
\begin{center}
\includegraphics[width=18cm]{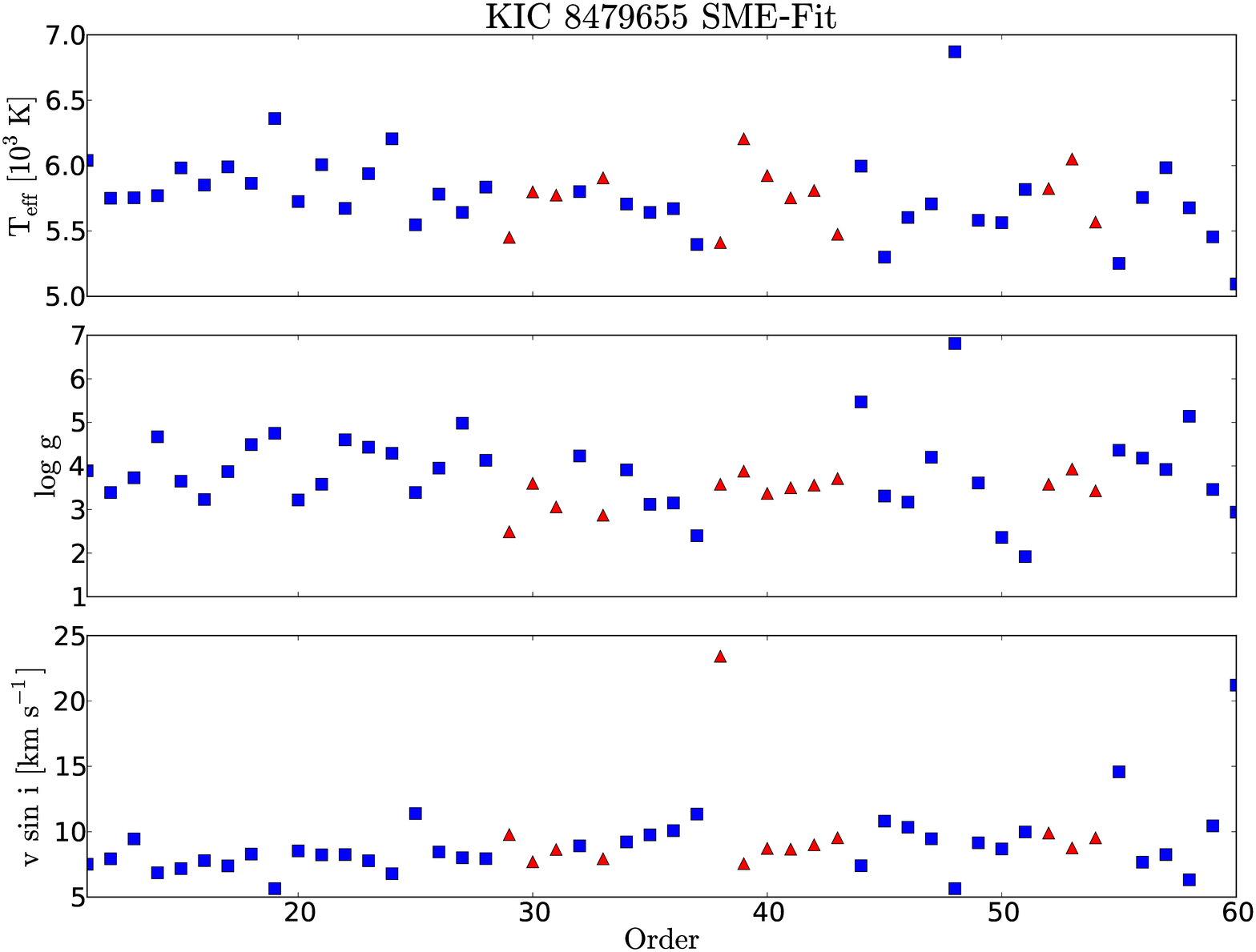}
\caption{\label{figure:KIC847} Best fitting parameters for all 
orders for KIC 8479655. The
orders marked with triangles are used for the fitting of the other objects.}
\end{center}
\end{figure*}

We based our SME fits on atomic data from 
VALD3\footnote{http://vald.inasan.ru/~vald3/php/vald.php?docpage=rationale.html}
(Kupka et al. \cite{VALD3}) and determined 
empirical oscillator strength
corrections to improve agreement between the synthetic and observed
spectrum of the Sun following the procedure described in
Valenti \& Fischer (\cite{valenti05}). 
As model input to SME, we used Kurucz model
atmospheres  (Kurucz \cite{Kurucz}). Furthermore, we set the micro turbulence
velocity to 0.85  km\,s$^{-1}$ and the macro turbulence to 3.0
km\,s$^{-1}$. The spectroscopic orders were fitted individually, and the low S/N 
regions at both ends of each order were clipped.
Wavelength regions showing cosmics were excluded from the analysis.
In a first fitting step, we determined the radial
velocity of each object. We then fitted 12 spectral orders
independently with T$_{\mathrm{eff}}$, log\,g, and v\,sin\,i as free
parameters for each order. The criteria for selecting those orders were
high S/N, avoidance of eventual contamination by chromosperic emission, and
avoidance of wavelength regions strongly affected by telluric lines. 
The wavelength ranges of these orders are listed 
in Table \ref{table:orders}.
For the three objects KIC 8479655, KIC 11073910,  and KIC 3626094, we
fitted all orders with  S/N $>$ 10 (a total number of 52 orders) to verify that
the fitting process with only 12 orders gives
parameters in agreement with fitting all usable orders. An example of the
determined parameters for each order can be found in Fig. \ref{figure:KIC847}.  
Only for the stars KIC 11972298 and KIC 11390058 did we choose 
different orders because its low S/N
made some of the orders unusable for the fitting. 
For each free parameter, we took the mean
of this fitting process as the final stellar parameters and its
standard deviation as a robust error estimate.

%______________________________________________________________
%
%_____________________________________________________________
%                                             Simple A&A Table
%_____________________________________________________________
%
\begin{table}
\caption{\label{table:orders} Start and end wavelengths of the orders 
used for determining stellar parameters.}
\centering
\begin{tabular}{rr}
\hline
\hline
Start &  End \\
$[\AA]$ & $[\AA]$ \\
\hline
5012 & 5066 \\
5057 & 5111 \\
5102 & 5157 \\
5196 & 5251 \\
5446 & 5504 \\
5499 & 5558 \\
5553 & 5612 \\
5608 & 5668 \\
5664 & 5724 \\
5721 & 5782 \\
6293 & 6361 \\
6364 & 6432 \\
\hline                                   %inserts single line
\end{tabular}
\end{table}

Furthermore, we conducted some upstream studies with SME to investigate
how our fitting results were sensitive to different methods of
analysis regarding the error estimation.  For KIC 8479655, KIC 3626094,
and KIC 11073910, we fitted our parameters T$_{\mathrm{eff}}$, log\,g,
and v\,sin\,i using the same pair of four wavelength intervals as
described in Valenti \& Fischer (\cite{valenti05}).  Our results show
that these methods are not more precise, and they lead to the same error
range as the method described above.  We also investigated
how sensitively our determined parameters depend on (i) the radial
velocity, which we varied  within twice the range of our error
estimates,  and (ii) variations in the macro turbulence
velocity. The latter is important, since we did not re-fit the 
stellar parameters with the correct macro turbulence velocity after 
obtaining the effective temperature. Both variations in the radial 
velocity and the macro turbulence velocity had a very weak effect 
on the derived stellar parameters, which was negligible in comparison 
to our error estimates. We repeated the analysis with [M/H] as
additional free parameter.    The metallicity analysis
was only possible for the three objects with highest S/N and slowest
rotation.

%______________________________________________________________
%
%_____________________________________________________________
%                                             A&A Figure
%_____________________________________________________________
%

For the PHOENIX model spectra fits, we used a grid with
T$_{\mathrm{eff}}$ ranging from 5000 to 6600 K in steps of 100 K and
log\,g ranging from 4.0 to 5.5 in steps of 0.5. During the $\chi^{2}$-fitting
process, we rotationally broadened the model spectra with v\,sin\,i
ranging from 3 to 27  km\,s$^{-1}$ in steps of 3 km\,s$^{-1}$, except for
KIC 9653110, where we varied v\,sin\,i from 30 to 100 km\,s$^{-1}$ in 
steps of 10 km\,s$^{-1}$.
We did not fit the metallicity with PHOENIX, but assumed solar
metallicity. We used 60 of the 80 orders for the fitting process and
again determined the final stellar parameters as the mean and its
standard deviation. 

The stellar parameters determined using both methods can be found in Table
\ref{table:parameters}.  Heliocentric velocity corrections were computed
using the IRAF `rvcorr' task. The parameters found by the SME fit with 
metallicity as additional free parameter can be found in 
Table \ref{table:metallicity}.  For the binary
star KIC 7264976, we used only PHOENIX spectra to determine the stellar
parameters of both components, which are shifted relatively to each
other by 12 km\,s$^{-1}$, since SME is not suited to fitting binary stars.

Generally, the PHOENIX fitting method and SME give similar results, which
usually agree within their respective errors. 
However, the PHOENIX fits systematically 
yield slightly higher effective temperatures.

%______________________________________________________________
%
%_____________________________________________________________
%                                             Wide A&A Table
%_____________________________________________________________
%
\begin{table*}
\caption{\label{table:parameters} Stellar parameters determined with 
SME and PHOENIX. }
\centering
\begin{tabular}[htbp]{cccccccccccc}
\hline
\hline
KIC      & S/N& v$_{\mathrm{r}}$ &  PHX T$_{\mathrm{eff}}$ & SME T$_{\mathrm{eff}}$& log\,g PHX & log\,g SME & v\,sin\,i PHX & v\,sin\,i SME\\
         &    &[km\,s$^{-1}$]  & [K]            & [K]            &               &               & [km\,s$^{-1}$] & [km\,s$^{-1}$]\\
\hline
 3626094 & 50 & -46 $\pm$ 1&  6000 $\pm$ 200 & 5900 $\pm$ 100 & 4.0 $\pm$ 0.5 & 4.2 $\pm$ 0.3 &  9 $\pm$ 3 &  6.1 $\pm$ 0.4 \\
 4742436 & 60 & -55 $\pm$ 1&  6100 $\pm$ 150 & 5945 $\pm$ 90  & 4.0 $\pm$ 0.5 & 4.3 $\pm$ 0.3 &  9 $\pm$ 3 &  7.0 $\pm$ 0.6 \\
 4831454 & 70 & -25 $\pm$ 1&  5600 $\pm$ 150 & 5530 $\pm$ 85  & 4.5 $\pm$ 0.5 & 4.6 $\pm$ 0.2 &  6 $\pm$ 3 &  5.5 $\pm$ 0.4 \\
 7264976A& 50 & -11 $\pm$ 2&  5750 $\pm$ 150 &                & 5.0 $\pm$ 0.5 &               & 18 $\pm$ 5 &                \\
 7264976B&    & -11 $\pm$ 2&  5100 $\pm$ 200 &                & 5.0 $\pm$ 0.5 &               &  9 $\pm$ 3 &                \\
 8479655 & 20 & -35 $\pm$ 1&  6200 $\pm$ 300 & 5670 $\pm$ 200 & 4.5 $\pm$ 0.5 & 3.7 $\pm$ 0.4 &  9 $\pm$ 3 &  9.0 $\pm$ 1.5 \\
 9653110 & 17 & -35 $\pm$ 3&  6000 $\pm$ 400 & 5700 $\pm$ 400 & 4.0 $\pm$ 0.5 & 4.5 $\pm$ 0.9 & 80 $\pm$ 10& 87.0 $\pm$ 8.9 \\
11073910 & 45 &  -1 $\pm$ 1&  6500 $\pm$ 150 & 6750 $\pm$ 400 & 4.5 $\pm$ 0.5 & 4.8 $\pm$ 0.5 & 18 $\pm$ 3 & 11.2 $\pm$ 1.3 \\
11390058 & 17 & -23 $\pm$ 3&  6000 $\pm$ 400 & 5740 $\pm$ 300 & 4.5 $\pm$ 0.5 & 4.5 $\pm$ 0.8 & 12 $\pm$ 3 &  9.9 $\pm$ 3.9 \\
11610797 & 50 & -14 $\pm$ 3&  5900 $\pm$ 200 & 5650 $\pm$ 150 & 4.0 $\pm$ 0.5 & 4.1 $\pm$ 0.6 & 24 $\pm$ 3 & 28.4 $\pm$ 1.1 \\
11764567 & 17 &  -1 $\pm$ 3&  6100 $\pm$ 300 & 5640 $\pm$ 240 & 4.5 $\pm$ 0.5 & 3.5 $\pm$ 0.8 & 21 $\pm$ 3 & 19.3 $\pm$ 2.5 \\ 
11972298 & 17 &  12 $\pm$ 3&  6100 $\pm$ 400 & 5500 $\pm$ 400 & 4.0 $\pm$ 0.5 & 3.7 $\pm$ 1.0 & 21 $\pm$ 6 & 17.6  $\pm$ 10.2\\
\hline
\end{tabular}

\end{table*}

%______________________________________________________________
%
%_____________________________________________________________
%                                             Simple A&A Table
%_____________________________________________________________
%
\begin{table}
\caption{\label{table:metallicity} Stellar parameters determined 
with SME including the metallicity as free parameter. }
\centering
\begin{tabular}[htbp]{ccccr}
\hline
\hline
KIC &  T$_{\mathrm{eff}}$&  log\,g & v\,sin\,i & [M/H]\\
       & [K] &  &  [km\,s$^{-1}$] & \\
\hline
3626094 & 5950 $\pm$200 & 4.3 $\pm$0.4 & 5.8 $\pm$0.3 & -0.01 $\pm$0.13\\
4742436 & 5850 $\pm$200 & 4.1 $\pm$0.4 & 6.7 $\pm$0.5 & -0.09 $\pm$0.18\\
4831454 & 5610 $\pm$120 & 4.8 $\pm$0.2 & 5.3 $\pm$0.4 &  0.03 $\pm$0.17\\
\hline
\end{tabular}
\end{table}

\subsection{Stellar activity}

We tried to infer the activity level of the stars from several
strong chromospheric lines.  We compared the best-fitting PHOENIX
spectrum to the observed spectrum to find excess line
emission or absorption for the strongest chromospheric lines in our
stellar spectra, namely H$\alpha$, \ion{Na}{i} D, 
\ion{Ca}{ii} at 8498, 8542, 8662~\AA, H$\beta$, H$\gamma$, and H$\delta$.

The best line indicators are H$\alpha$ and the \ion{Ca}{ii} infrared triplet, 
since these are strong, unblended lines with a well defined continuum level.
The \ion{Na}{i} D lines are significantly affected by the presence of 
strong telluric lines in their vicinity, 
and the higher Balmer lines suffer from both the lack of a well defined continuum 
and the declining flux in the blue region of the spectrum.

\begin{figure*}
\begin{center}
\includegraphics[width=18cm,height=18.5cm,bbllx=65,bblly=1,bburx=558,bbury=368]{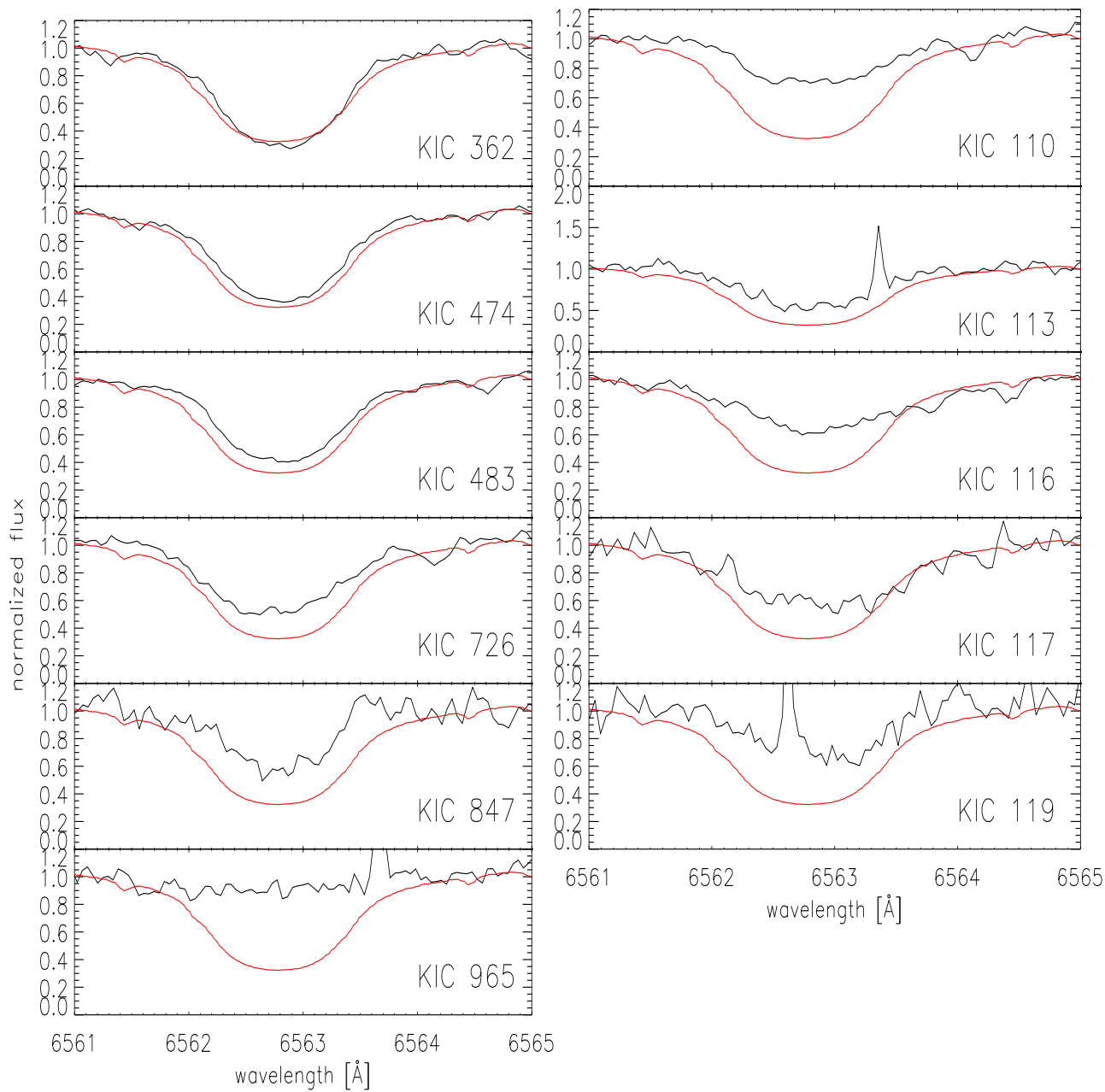}
\caption{
  \label{halpha} H$\alpha$ line for the sample stars and the inactive
  star HD 115383. 
  The asymmetries in the red wing of
  the line in KIC 8479655 (emission feature at about 6563.5 \AA) and 
  KIC 11610797 (broad additional absorption feature between about 
  6563 and 6564.5 \AA) are clearly visible.
  KIC 9653110 and KIC 11972298 both have a cosmic in the H$\alpha$ line. 
  % KIC 11390058 is not shown here, but its spectrum is very similar 
  % to KIC 11764567 and KIC 8479655.  
  }
\end{center}
\end{figure*}

We also compared these chromospheric lines to an observed spectrum of
an inactive G0 dwarf star, HD 115383 (59 Vir), from the UVES spectral
atlas\footnote{The UVES spectral atlas is available at
  https://www.eso.org/sci/observing/tools/uvespop/field\_stars\_uptonow.html}.
This yielded results that are similar to the comparison to the best-fitting 
PHOENIX spectrum. As an example of chromospheric activity, 
in Fig. \ref{halpha} we show the H$\alpha$ line 
for the stars in our sample, as well as for the reference star
HD 115383.
The reference star is not rotationally broadened,
since for the H$\alpha$ line, rotationally broadening plays no 
significant role in the velocity regime considered here.  

In Sect. \ref{individual}
we note for each star which chromospheric lines are filled in. We
measured absolute H$\alpha$ EW, but these are
hampered by the fact that in many cases, the Voigt profile used results
in a poor fit. We therefore computed the EW by integrating
the residual line after subtracting a quiescent reference star, again 
using the spectrum of HD 115383 for the latter. We estimate our errors to
be about 20 - 30 m\AA\, for the stars with S/N of 45 and higher,
and about 50 m\AA\,
for the low S/N stars because of the larger uncertainty in the normalisation.
For the EW computation, the effect of cosmics has been removed by 
replacing them with the mean values of neighbouring pixels, but to show their
location we did not remove them in Fig. \ref{halpha}.

This method has also been used by Froehlich et al. (\cite{froehlich}) and
Soderblom et al. (\cite{soderblompl}), 
who give a subtracted EW(H$\alpha$)
between 100 and 500 m\AA\, for early G
stars at the age of the Pleiades.
Our EW values, given in in Table
\ref{table:2}, match these values, with the notable exception of KIC 3626094.
 
Two of the stars (KIC 11610797 and KIC 8479655) show asymmetries 
in their H$\alpha$ lines that significantly affect the EW(H$\alpha$) 
measurements. We discuss this in more detail in Sect. \ref{individual}.

%______________________________________________________________
%
%_____________________________________________________________
%                                             A&A Figure
%_____________________________________________________________
%

\subsection{Lithium}

Lithium  is rapidly burned at the bottom of the convective 
zone in low-mass stars, and therefore its line at 6707 \AA\, 
is often used as an age indicator.
While there is substantial scatter at a given T$_{\mathrm{eff}}$ even within open
clusters, that is to say among coeval stars (e.g. Soderblom et al. \cite{soderblom93}),
this scatter occurs downward from an upper limit in the 
(EW$_{\ion{Li}{i}}$ -- T$_{\mathrm{eff}}$)
diagram. In particular, the upper limit for the Pleiades, which is defined well 
by a large number of measurements, is a robust benchmark in the sense
that no older stars are located {\it above} it (Wichmann \cite{wichmann00}).
The Pleiades have an age of about 125 Myr (Tognelli et al. \cite{tognelli}).

\begin{figure*}
\begin{center}
\includegraphics[width=18cm]{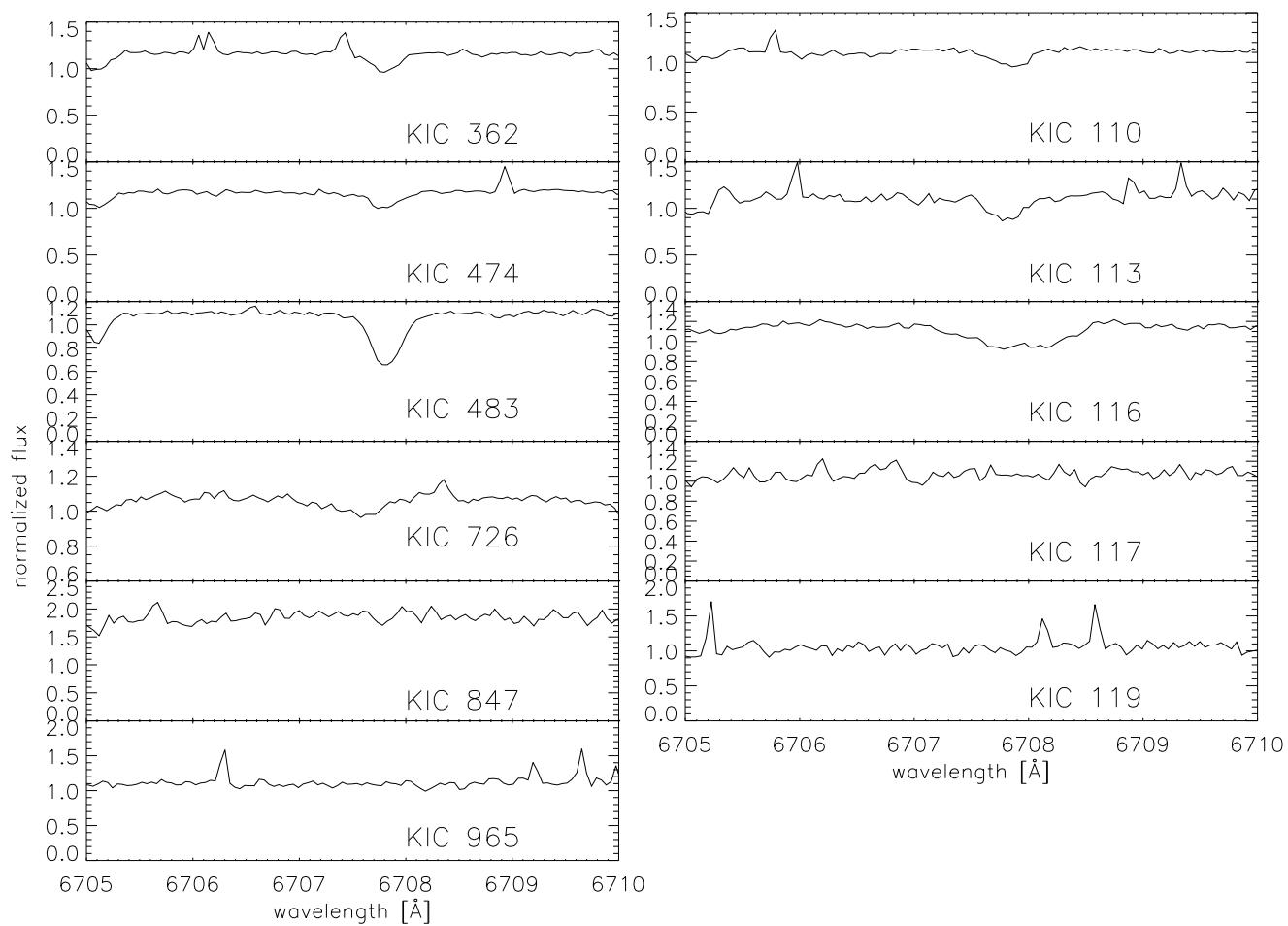}
\caption{\label{figure:lithium} Spectra of our programme stars showing
the wavelength region around the \ion{Li}{i} line at 6707 \AA.}
\end{center}
\end{figure*}

The EW of \ion{Li}{i} at 6707 \AA\, for our programme stars 
has been measured with IRAF 
using the {\sc splot} task. For each star, three independent measurements 
have been averaged to form the final result. The main source of error
is the estimation of the local (pseudo-)continuum, which leads to an r.m.s.
error of $\approx$\,5\,m{\AA}. The spectral region around the 
\ion{Li}{i}\,6707.44\,\AA\, line is shown in Fig. \ref{figure:lithium}, and
the measured results are given in 
Table \ref{table:2}, along with the {\it expected} contribution from the
blend with the \ion{Fe}{i}\,6707.44\,\AA\, line 
(Favata et al. \cite{favata93}, Fig.\,1).

%______________________________________________________________
%
%_____________________________________________________________
%                                             Simple A&A Table
%_____________________________________________________________
%
\begin{table}
\caption{Measured EW for individual lines.}             % title of Table
\label{table:2}      % is used to refer this table in the text
\centering                          % used for centering table
\begin{tabular}{r r r r }        % centered columns (4 columns)
\hline\hline                 % inserts double horizontal lines
    &            &             &   subtracted   \\
KIC & EW$_{Li}$   & EW$_{Fe}$    &    H$\alpha$ EW      \\  % Fe 6707.44
    & [m\AA]     & [m\AA]      &      [m\AA]        \\ % (log T)
\hline
   3626094 &  56 &  4          &  50       \\ % (3.77) PL
   4742436 &  67 &  8          &  120      \\ % (3.75) PL
   4831454 & 154 & 13          &  220      \\ % (3.72) PL
   7264976 &  76 & 16          &  330      \\ % (3.71) PL
   8479655 &   - & 17          &  430      \\ % (3.71)
   9653110 & ... & ...         &  550      \\
  11073910 &  66 & 12          &  460      \\ % (3.73) PL
  11390058 &  93 &  4          &  340      \\ % (3.78) ZAMS
  11610797 & 210 &  4          &  310      \\ % (3.77) ZAMS
  11764567 &   - & 15          &  230      \\
  11972298 & ... & ...         &  430      \\
\hline                                   %inserts single line
\end{tabular}
\end{table}

%______________________________________________________________
%
%_____________________________________________________________
%                                             A&A Figure
%_____________________________________________________________
%

\subsection{\textit{Kepler} light curves}
\begin{figure*}
\begin{center}
\includegraphics[width=18cm]{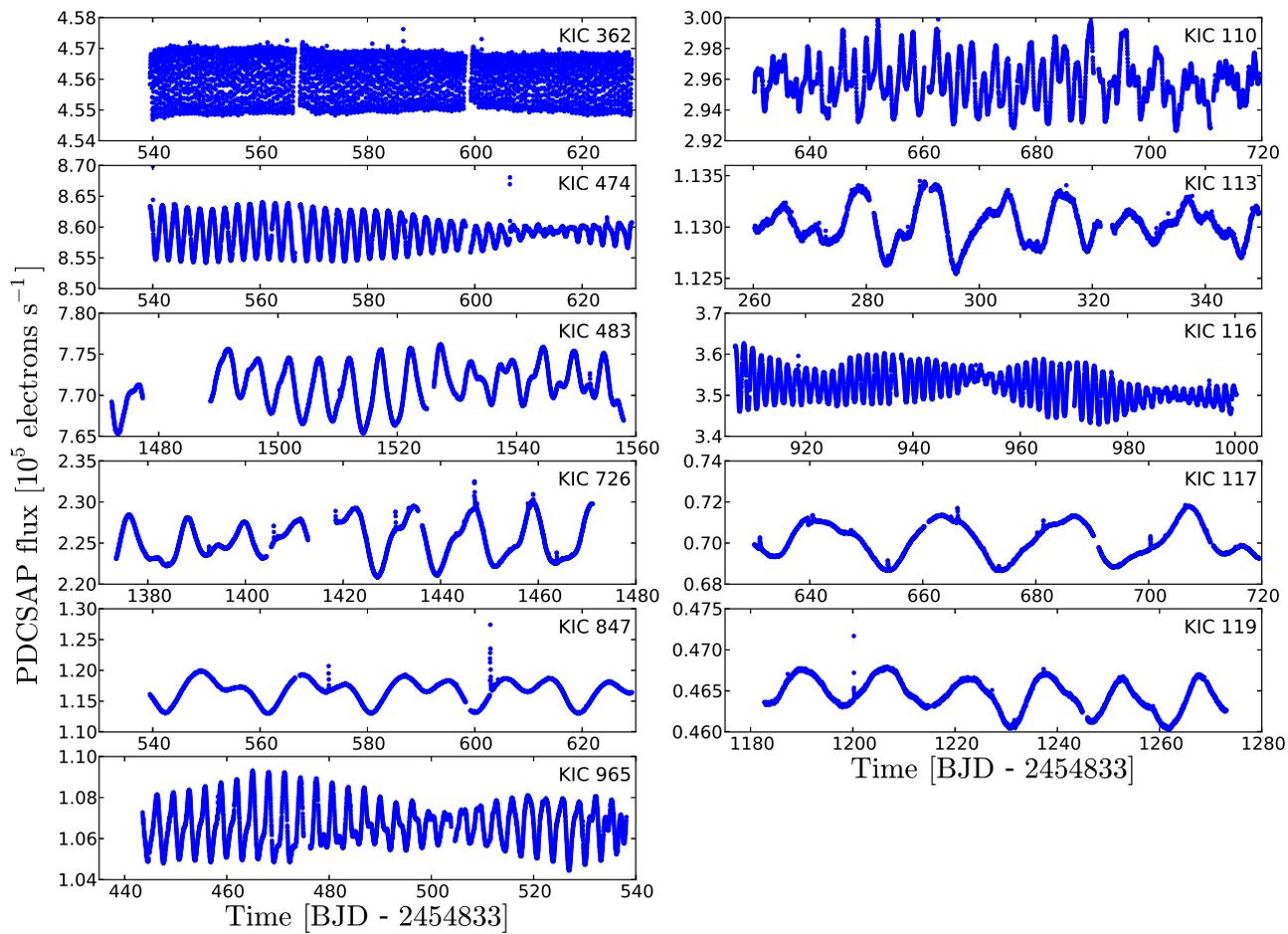}
\caption{\label{kepler_lightcurves} \textit{Kepler} light curves for the 
target stars. 
%Left Column from top to bottom: 
%KIC 3626094 (quarter 6), KIC 4742436 (quarter 6), KIC 4831454 (quarter 16), 
%KIC 7264976 (quarter 15) KIC 11390058 (quarter 3), 
%and KIC 11610797 (quarter 10). 
%Right column from top to bottom:
%KIC 8479655 (quarter 6), KIC 9653110 (quarter 5), KIC 11073910 (quarter 7), 
%KIC 11764567 (quarter 7), and KIC 11972298 (quarter 13).}
}
\end{center}
\end{figure*}

In Fig. \ref{kepler_lightcurves} we show the \textit{Kepler} light 
curves of our target stars as taken from the NASA exoplanet 
archive\footnote{The web page can be found under:
http://exoplanetarchive.ipac.caltech.edu/index.html}. 
We used the Lomb-Scarge periodogram analysis provided
as a service by the NASA exoplanet archive to search for periods in the
light curves.
Generally all stars show more than
one significant peak in the periodogram mostly due to aliasing. 

The light curve patterns of the stars in our sample can be divided 
into three groups. The first and largest group shows a well defined 
periodic behaviour, although there seem to be secondary maxima, some of 
which are drifting slowly 
with respect to the main maximum (KIC 3626094, KIC 4831454, KIC 7264976, 
KIC 8479655, and KIC 11764567). 
The second group of light curve patterns is similar to the first but 
additionally shows a distinct beating pattern in its light curve 
(KIC 9653110, KIC 4742436, and KIC 11610797). The beat period 
is not well defined for any
of the three stars and not found unambiguously in the 
Lomb-Scarge periodogram for individual Kepler quarters.
The last group consists of KIC 11073910, KIC 11390058, and KIC 11972298, 
where the light curve shows no distinct pattern, or a periodic 
behaviour is found only in some quarters.
For these stars it is not clear to us what causes the complicated 
light curve (see also Sect. \ref{individual}).  

Similar beat patterns have been observed, as seen on for instance \textit{CoRoT}-2 by
Huber et al.  \cite{Huber}, who present two different 
interpretations for the beat pattern. 
First it can be explained by differential rotation
of at least three spots or active regions.  
Alternatively, it can be
explained by a flip-flop effect: There are two active regions on
different hemispheres of the star, with one dominating the light curve. During
the minimum of the beat pattern, the other active region rapidly becomes the
dominant one.  
This kind of behaviour has also been observed in
the pre-\textit{Kepler} era in single stars, using temperature mapping
techniques (see e.\,g. Korhonen et al. \cite{Korhonen}).

Most of the stars in our sample show 
periods that are at least double-peaked, with one peak
moving slowly with respect to the other. This slow movement of the two peaks
relative to each other identifies the feature as being caused by different spots
on the surface of the star. Nevertheless, a double star system cannot be excluded as the origin
of this light curve pattern, if both stars are active
and are not synchronised.
Therefore, all stars in our sample seem to be active, in the sense that they
have different spot groups on their surface, manifesting in complicated but
periodic light curves. The relative movement of the peaks may be caused by
differential rotation of the star, but it may also be caused by spot evolution
(see Huber et al. \cite{Huber}).

\section{Individual stars}\label{individual}

\subsection{KIC 3626094}

KIC 3626094 (2MASS J18591491+3845462, TYC 3119-1230-1) has the shortest
rotation period within our sample (0.724\,d). The spectrum shows
no obvious signs of activity, whether compared to the PHOENIX best fit spectrum
or to the inactive G0 star. From the chromospheric point of view,
KIC 3626094 is the most inactive star of the sample, which agrees well with
it not being detected as a ROSAT source.
Nevertheless, Li $\lambda\,6707$ 
is clearly detected, albeit with an EW below the Pleiades upper limit.

Ultra fast rotators with periods well below one day among G-type stars
are typical of young open clusters up to the age of the Pleiades
(c.f. Barnes \& Sofia \cite{barnes96}). We therefore conclude that
KIC 3626094, with its T$_{\mathrm{eff}}$ of $\approx 5800-6000$ K, is a young
G0 - G2 star with an age similar to, or less than, the Pleiades
(i.e. $\lesssim 125$\,Myr).

\subsection{KIC 4742436}

KIC 4742436 (TYC 3138-950-1) can be identified with the ROSAT X-ray source 
1RXS J192149.3+395017 (offset 10.4\arcsec), which has a count rate
of 0.0121 ct/s and a positional error of 14\arcsec 
(Voges et al. \cite{voges00}). 
Its rotation period of 2.3 days is well below what is observed for G-type stars
in the Hyades cluster (see Fig. 1 in Barnes \& Sofia \cite{barnes96}),
and the \ion{Li}{i} $\lambda\,6707$ line is clearly detected. But with regard 
to the chromospheric emission lines, this star is one of the most inactive 
of the sample.
KIC 4742436 therefore appears to be a modestly active young G-type 
star (younger than the Hyades cluster at $\approx$ 600\,Myr).

\subsection{KIC 4831454}

KIC 4831454 (TYC 3138-1050-1) may be the ROSAT X-ray source 
1RXS J192200.0+395957 (offset 22.5\arcsec), which is listed with a 
positional error of 17\arcsec and a count rate of 
0.0166 ct/s. Another sign of its activity is the observed filling-in of
\ion{Ca}{ii} lines, although the H$\alpha$ line is not as strongly filled
as for other stars in the sample.

With a period of 5.19 days, the star rotates faster than Hyades stars
of comparable B-V (0.74 as measured by TYCHO, Hog et al. \cite{tyc}).
It also shows a relatively strong \ion{Li}{i} $\lambda\,6707$ line,
and we conclude that, like KIC 4742436, this star is probably younger
than the Hyades.

\subsection{KIC 7264976}

The spectrum of this star (TYC 3128-361-1) shows a double line profile, 
indicating that it is a binary star. However, we can detect 
no measurable velocity
variation during our observation run. Also the 
spectrum obtained during our 2013 run shows no measurable line shifts.

Since it seems very unlikely to have two unrelated stars of similar magnitude
at the same position in the sky (within $\approx$ 1\arcsec), we consider it 
more likely that the star is a binary with a long period of $\sim$ 270 days
and that we observed in both runs near quadrature when the variation in
the radial velocities would be slowest. If this assumption is correct, then
the star is a relatively wide binary ($\approx$ 1 AU). The binarity would
not explain the observed flaring activity as active binaries of the RS CVn type
are tight binaries where the stars spin fast because of tidal synchronisation.

The \ion{Li}{i} $\lambda\,6707$ line is clearly detected, although the
period of 12.7 days seems to hint at an age in excess of the Hyades cluster.
However, since the star is a binary, it is not clear which of the two
stars is responsible for the photometric period, and although the error
is large, the v\,$\sin{i}$ of KIC 7264976A seems to be at odds with the
period.

Chromospheric activity is seen in some filling in the H$\alpha$ line and
in the \ion{Ca}{ii} infrared triplet.

\subsection{KIC 8479655}

KIC 8479655 (2MASS J18574324+4435558) does not show the 
\ion{Li}{i} $\lambda\,6707$ line in its spectrum, although the
filling-in of the H$\alpha$ and \ion{Ca}{ii}  lines indicate
some chromospheric activity. The H$\alpha$ line shows some asymmetry in the
red wing, with an emission component (see Fig. \ref{halpha}). However, this
may be caused by noise, since it is the star with one of the lowest 
S/N in the sample.
Therefore, the measured H$\alpha$ EW is presumably too small.

Based on the non-detection of \ion{Li}{i} $\lambda\,6707$ and the photometric
period, this star seems to be significantly older than the other ones in
our sample, and it is not clear why it shows
chromospheric activity and strong flares.
It has a long
photometric period of 19.3 days, which seems to be inconsistent with the
measured v\,$\sin{i}$ of 9\,$\pm$\,3 km\,s$^{-1}$ even if the temperature
were significantly in error, since for an F-G main sequence
star, we would expect a radius in the range of 0.85 to 1.3 R$_{\sun}$ 
(Zombeck \cite{zombeck}),hence a rotational velocity of $\approx$ 3\,--\,2 km\,s$^{-1}$.
The minimum radius of KIC 8479655 is $3.4\pm1.2$\,R$_{\sun}$.
Our spectroscopically determined log\,g value is inconsistent with this 
star being a giant, although it is compatible with a subgiant. 
We do not find significant radial velocity variations
during the timespan of our 2012 observations, but we cannot rule out that
the star may be a wide binary, with a brighter component dominating the
spectrum and a fainter component dominating the light curve variations.

\subsection{KIC 9653110}

KIC 9653110 is a fast-rotating star, but it was not detected in the 
ROSAT All-Sky Survey. The spectrum has a low S/N that, in combination with
the rotational broadening, makes it impossible to determine whether the
\ion{Li}{i} $\lambda\,6707$ line is present or not.

With respect to rotation and temperature, the star appears to be similar
to KIC 4831454 and KIC 4742436, and we conclude that like these two stars,
it is probably a young star, at least younger than the Hyades.
Its chromospheric activity manifests in all investigated lines, they 
are all filled up. It is the most active star of the sample.

We note that a stellar radius of at least of $5.0\pm0.6$ solar radii is 
required to reconcile the rotation period with the v\,$\sin{i}$ value.
While the log\,g may be compatible with a subgiant, we find it very 
unlikely that a subgiant would have a v\,$\sin{i}$ of approximately 
80\,km\,s$^{-1}$. Therefore we presume that the star is either a
pre-main sequence star, or the rotation period is due
to spots on a companion star that is too faint to show up in the spectrum.

\subsection{KIC 11073910}

KIC 11073910 (TYC 3545-1049-1) is a fast rotator with a clearly detected
\ion{Li}{i} $\lambda\,6707$ line, it shows chromospheric activity
in all investigated lines, and is the second most active star in the sample.
There is a ROSAT X-ray source (1RXS J190429.1+483725) at an offset
of 33.2\arcsec, listed with a positional error of 23\arcsec and a count
rate of 0.0267 ct/s.

We conclude that this is a young star with an age lower or equal to the Hyades
cluster (based on the observed period). Notably, the T$_{\mathrm{eff}}$ listed 
in the {\textit Kepler} input catalogue is significantly lower than the 
one we determine from our spectra.

Also, the \textit{Kepler} light curve of this object is different from the
other light curves. Though a period can be found using the periodogram (and
can also be established by inspecting the light curve by eye), the light curve
shows a lot of additional flickering, the cause of which is not clear. 
Possible explanations are a multitude of active regions, minor flares, or 
pulsations. If our inferred stellar parameters are
correct, the star is an early-to-mid F star and therefore close to the 
borderline between stars with and without convection zones. 
This leaves some doubt as to whether the detected period is
really caused by stellar spots. On the other hand, the star has filled in
chromospheric lines and seems to be chromospherically active.

\subsection{KIC 11390058}

All chromospheric lines are filled in to some degree and the star is 
among the more active ones in the sample. 
Nevertheless, no ROSAT source was found within a search radius of 120\arcsec. The \textit{Kepler} light curve shows no conclusive behaviour. Some quarters
show a periodic pattern of about 12 days as shown in 
Fig. \ref{kepler_lightcurves}. But this period is not found for every 
quarter. Some of these other quarters exhibit different periods, but some 
also look very irregular without a clear periodic behaviour. We attribute 
this behaviour to changing spot patterns.

Although the light curve hints at a slow rotation period, the spectrum clearly 
shows the \ion{Li}{i} $\lambda\,6707$ line, with an EW somewhat less 
than the Pleiades upper limit, indicating that this is a young star.

\subsection{KIC 11610797}

KIC 11610797 (TYC 3551-1852-1) shows a \ion{Li}{i} $\lambda\,6707$ line 
with an EW clearly in excess of the Pleiades upper limit, indicating that it
is significantly younger than the Pleiades cluster (125\,Myr). It is also
a rapid rotator and shows clear indications of chromospheric activity
in most of the lines we investigated. 
We do not detect any filling-in of the H$\gamma$ and H$\delta$ lines, 
which may be caused  by the rotational 
broadening of the lines, the low S/N, and problems in 
determining a continuum in this wavelength region. Also this star shows
an asymmetry in the red wing of the H$\alpha$ line, with additional absorption.
This asymmetry seems to be real and not caused by noise. Unfortunately, the
additional absorption component is very broad, extending from 
about 6563 to 6564.5 \AA\,
and not deep enough to be fitted with an additional Gaussian or Voigt profile.

This is most probably a very young star, and as with several other stars of
our sample, there is an X-ray source (1RXS J192737.8+493949) detected nearby
(offset 27.5\arcsec, listed with a positional error of 27\arcsec and a
count rate of 0.0186 ct/s). 

\subsection{KIC 11764567}

There is no ROSAT source within 120\arcsec of KIC 11764567. The rotational
period of 20.5\,d and the non-detection of \ion{Li}{i} $\lambda\,6707$ indicate 
that this is probably an old star. 
However, just as in the case of 
KIC 8479655, the v\,$\sin{i}$
is much too large for a main sequence star of this rotation period. 
The minimum radius is $8\pm1.2$\,R$_{\sun}$. We have
four spectra of this star (three from Aug 22, 2012, one from Aug 17, 2012), but
the S/N of the individual spectra is too low to check for line shifts. Given
the discrepant values of log\,g determined by both methods used and not detecting the Li line, the star may either be evolved or 
be a binary. However, the v\,$\sin{i}$ seems too high for an evolved star.

All measured chromospheric lines show some filling-in. Therefore, we 
consider the star to be active.

\subsection{KIC 11972298}

The spectrum of KIC 11972298 is noisy and nearly all chromospheric lines 
are hampered by cosmics. Nevertheless, all investigated chromospheric 
lines are filled in, and the star is amongst the most
active stars of the sample, from the chromospheric point of view.
There is no ROSAT source within 120\arcsec of KIC 11972298. The noise
level meant we could not perform EW measurements of the 
\ion{Li}{i} $\lambda 6707$ line.

The \textit{Kepler} light curve exhibits regular 
periodic behaviour in some quarters,
but in other quarters the light curve seems to show chaotic
behaviour, though the periodogram finds significant periods for every quarter.
Many quarters show a period of about eight days, some a period of 14 to 16
days. Also other significant periods are found in individual quarters.
It is somewhat unclear, whether the eight-day period is an alias of the 14- to 16-day
period. The period for quarter 13 shown in Fig. \ref{kepler_lightcurves} 
is 15.6 days.
It remains unclear what causes this irregular behaviour. A possible
explanation is quickly evolving spots, but there are other possibilities, such as pulsations.

As with some other stars in our sample, the v\,$\sin{i}$
is much too large for a main sequence star of this rotation period.
The minimum radius is $5.6\pm3.5$\,R$_{\sun}$. Again, the star might either
have a faint companion that is responsible for the photometric variability
or it is not a main sequence star. In the latter case, since, we cannot rule 
out the presence
of a Li line it may either be an evolved star or a pre-main sequence star.

\section{Discussion}

Flares on low-mass main sequence stars are caused by solar-type magnetic 
activity, and it is well known that this activity depends chiefly on rotation.
Hence, the most active stars are those that are fast rotators, 
either because they are young or because they are in close binary systems.

Young stars have spun up during contraction towards the main sequence, but have
not yet had time to spin down by magnetic braking, i.e. the interaction of 
their magnetic fields with the wind. Close binary stars can be fast rotators
irrespective of their age owing to synchronisation of their rotational and 
orbital periods caused by tidal forces.
On the other hand, even slowly rotating solar-type stars can occasionally
show relatively strong activity, as shown, for instance, by the famous white-light 
Carrington flare on the Sun (Carrington \cite{carrington}).

{\it Kepler} has provided an unprecedented data set of photometric observations.
Never before has such a large sample of stars been observed at high cadence for
a long period of time, and that makes it difficult to place the results in
context. For example, the observed super flares might be typical of active young stars,
but there is no young open cluster in the {\it Kepler} field to confirm this.

Maehara et al. (\cite{maehara}) divide their sample into fast 
(period $\leq$\,10d) and slow rotators (period $>$\,10d), 
with the former presumably young, while the latter are termed ``Sun-like'' by them. 
Our spectroscopic observations confirm that the fast-rotating stars
indeed show \ion{Li}{i} $\lambda\,6707$ absorption lines, indicative of
their youth. Several of them can also be identified with 
X-ray sources from the ROSAT All-Sky Survey.

Of the five slowly rotating stars in our sample, two 
(KIC 7264976 and KIC 11390058) also show \ion{Li}{i} $\lambda\,6707$ 
absorption and might be young objects. 
We note that for KIC 7264976, it is not known which of the 
two components of this 
binary star causes the photometric period. 

The three other slow rotators (KIC 8479655, KIC 11764567, and
KIC 11972298), as well as one fast rotator (KIC 9653110) 
show a puzzling discrepancy between the
observed photometric rotation period and the measured v\,$\sin{i}$ that 
requires a minimum radius exceeding the
expected value (for a main sequence star) by a factor of 2 -- 8 for each of these stars.
We hypothesise
that these stars are either evolved (or pre-main sequence, when the Li line
cannot be measured), or they are binaries where the photometric
period originates in the less massive component, while the spectrum is 
dominated by the more massive one. We note that within errors, the measured 
values of log\,g might be compatible with these stars being subgiants, which 
would relax the tension between rotation period and v\,$\sin{i}$.
However, because of the magnetic braking experienced during their main sequence
life, stars in this mass range would be slow rotators at the end of their 
main sequence life and become even slower when evolving towards the giant
branch. We do not consider it likely that stars with v\,$\sin{i}$ significantly
in excess of the Sun are evolved subgiants.

In conclusion we find no general picture of the analyzed super-flare stars. 
Though many of them show indicators of youth that would be a straightforward
explanation of the exhibited super-flares, for others the situation is not so
clear or even puzzling. Besides the individual discrepancies discussed above, 
two of the stars showed very little chromospheric activity
during our observations, several stars - in particular more than half of 
the slow rotators - apparently have minimum radii in excess of main sequence 
stars, and some stars cannot
be associated with ROSAT sources. Therefore, more observations of super-flare
stars are needed in order to discover some common element or to confirm that
super flares are possible on a wider variety of stars.

\begin{acknowledgements}

  Part of this work was supported by the German
  \emph{Deut\-sche For\-schungs\-ge\-mein\-schaft, DFG\/} project
  numbers Wi~1669/2--1, WO~1645/3--1, and WO~1645/4--1. 
  We thank the staff at Calar Alto observatory 
  for their help and support.
  This research made use of the NASA Exoplanet Archive, which is 
  operated by the California Institute of Technology, under contract with 
  the National Aeronautics and Space Administration under the Exoplanet 
  Exploration Program.
\end{acknowledgements}

%-------------------------------------------------------------------

\end{document}